\documentclass[11pt,twoside]{article}
\usepackage{./asp2014}
\usepackage{amsmath}
\usepackage{array}
\usepackage{txfonts}
\usepackage{ifthen}
\usepackage{lscape}
\usepackage{index}
\usepackage{graphicx,amssymb}
\usepackage{wrapfig}
\usepackage{chapterbib}

\aspSuppressVolSlug
\resetcounters

\begin{document}

\title{Detection of binary and multiple systems among rapidly rotating K and M dwarf stars from Kepler data}
\author{Katalin~Ol\'ah,$^1$, Saul Rappaport$^2$, and Matthew Joss$^2$
\affil{$^1$Konkoly Observatory, MTA CsFK CsI, Budapest, Hungary; \email{olah@konkoly.hu}}
\affil{$^2$MIT, MA, USA; \email{sar@mit.edu};\email{mattjoss@mit.edu}}}

\paperauthor{Katalin Ol\'ah}{olah@konkoly.hu}{ORCID_Or_Blank}{Konkoly Observatory}{}{Budapest}{}{1121}{Hungary}
\paperauthor{Saul Rappaport}{sar@mit.edu}{ORCID_Or_Blank}{Kavli Institute}{Author2 Department}{City}{MA}{Postal Code}{USA}
\paperauthor{Matthew Joss}{mattjoss@mit.edu}{ORCID_Or_Blank}{Kavli Institute}{Author2 Department}{City}{MA}{Postal Code}{USA}

\begin{abstract}
From an examination of $\sim$18,000 Kepler light curves of K- and M-stars we find some 500 which exhibit rotational periods of less than 2 days. Among such stars, approximately 50 show two or more incommensurate periodicities.   We discuss the tools that allow us to differentiate between rotational modulation and other types of light variations, e.g., due to pulsations or binary modulations.  We find that these multiple periodicities are independent of each other and likely belong to different, but physically bound, stars.  This scenario was checked directly by UKIRT and adaptive optics imaging, time-resolved Fourier transforms, and pixel-level analysis of the data. Our result is potentially important for discovering young multiple stellar systems among rapidly rotating K- and M-dwarfs.
\end{abstract}

\section{The program and methods}
To study stellar activity at the lower end of the main sequence, all dwarf K and M type stars from the {\em Kepler} database were selected. From those, using a simple Fourier-transform search, we found a total of about 500 objects which exhibit light variations with short period(s) of less than 2 days. These objects were further studied to identify those stars which have spots on their surfaces, and consequently show rotational modulation in their observed light curves. We eliminated targets with obvious binary modulation, including those in the {\em Kepler} binary catalog.  We also concluded that K-M dwarf stars could not pulsate with the periods and amplitudes observed ($\sim$6 hr to 2 days, and amplitudes of a few percent).  More details about target selection and validation are found in \cite{saul}. Fig.~1 shows a typical example: KIC~8416220, which exhibits large flares (left panel); two close periods in the Fourier transform (right panel); and short period light variations, including `beats' (middle panel). 

\articlefigurethree{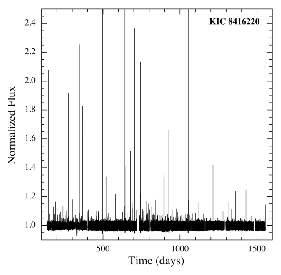}{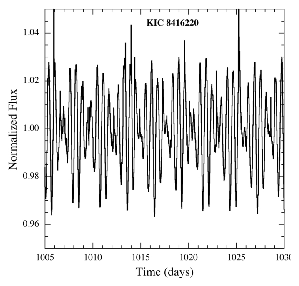}{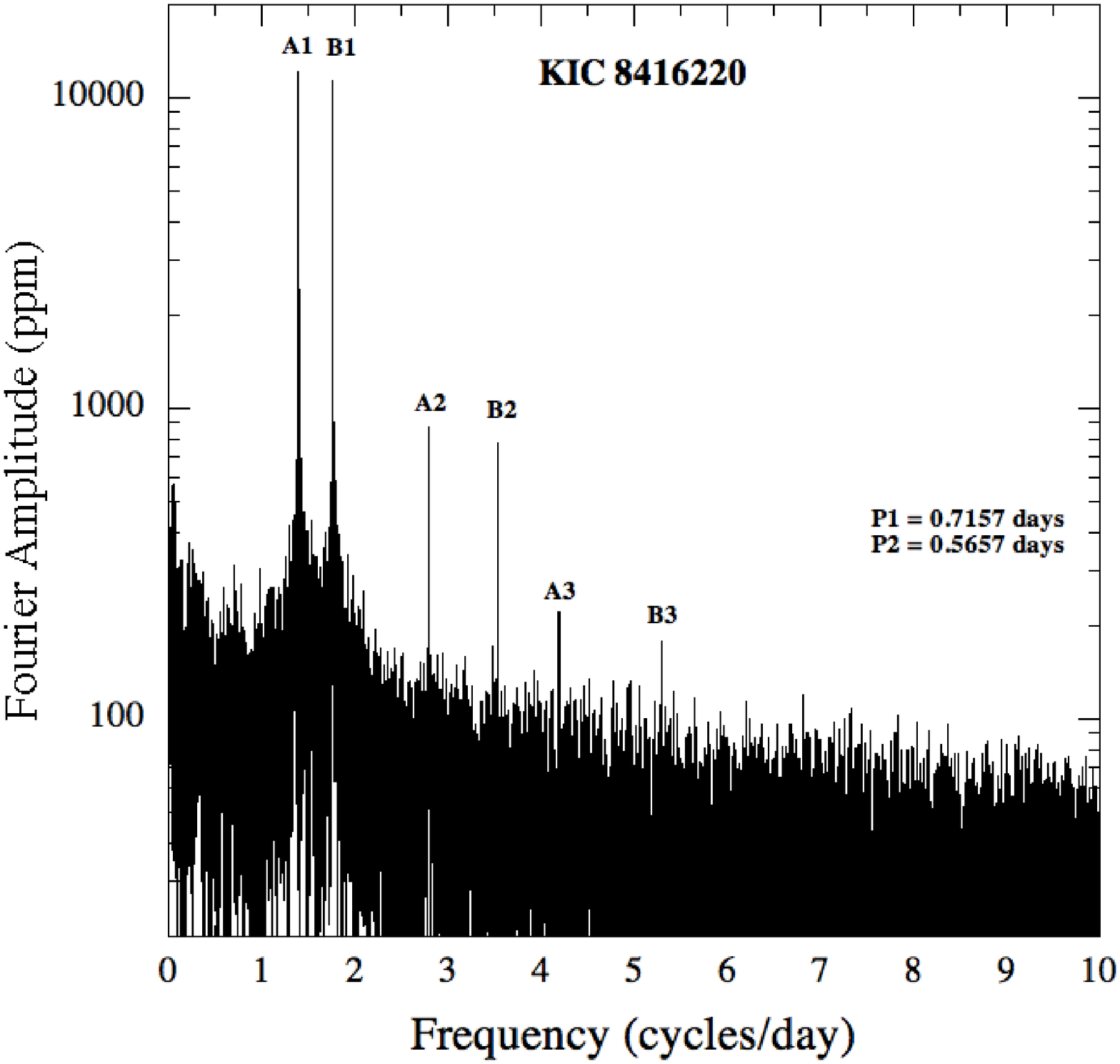}{ex_fig1_triple}{Light variations and the periods of the rotational modulation of KIC~8416220. \emph{Left:} full dataset showing large flares. \emph{Middle:} a small portion of the data with typical rotational modulation, including `beats'. \emph{Right:} Fourier spectrum revealing two close periods in the data. }

The stability/variability of the light curves were further studied by tracking the phases of the base frequencies and their first harmonics. In the case of stable light curves (e.g., due to binary eclipses) the phases behave simply and systematically with time. However, for modulation due to rotating starspots that may be differentially rotating, and can grow and decay with time, the phases of the base period and its first harmonic typically show erratic behaviour. Fig.~2 shows an example for each of these cases. In the left panel, phase tracking of the contact binary KIC~5128972 is shown: the phase behaviour of the half period (i.e., due to the ellipsoidal light variation) reflects the orbit of the close binary with respect to an inferred third body in the system. The right panel shows the typical erratic phase-tracking behavior of a spotted star, KIC~10710753.

We followed the time behaviour of the light curves with short-term Fourier transforms (STFT, "sonograms") which describe how the amplitudes and periods change with time. This approach gives good resolution both in time and in frequency. In these cases, data segments of about 30-day duration are Fourier transformed, and the centers of the data intervals are shifted by 2 days between adjacent points in the sequence. The resulting amplitudes are then plotted as as function of frequency in the vertical direction, with time running along the horizontal axis. Details of this method are given in \cite{ol_ko}.

\articlefiguretwo{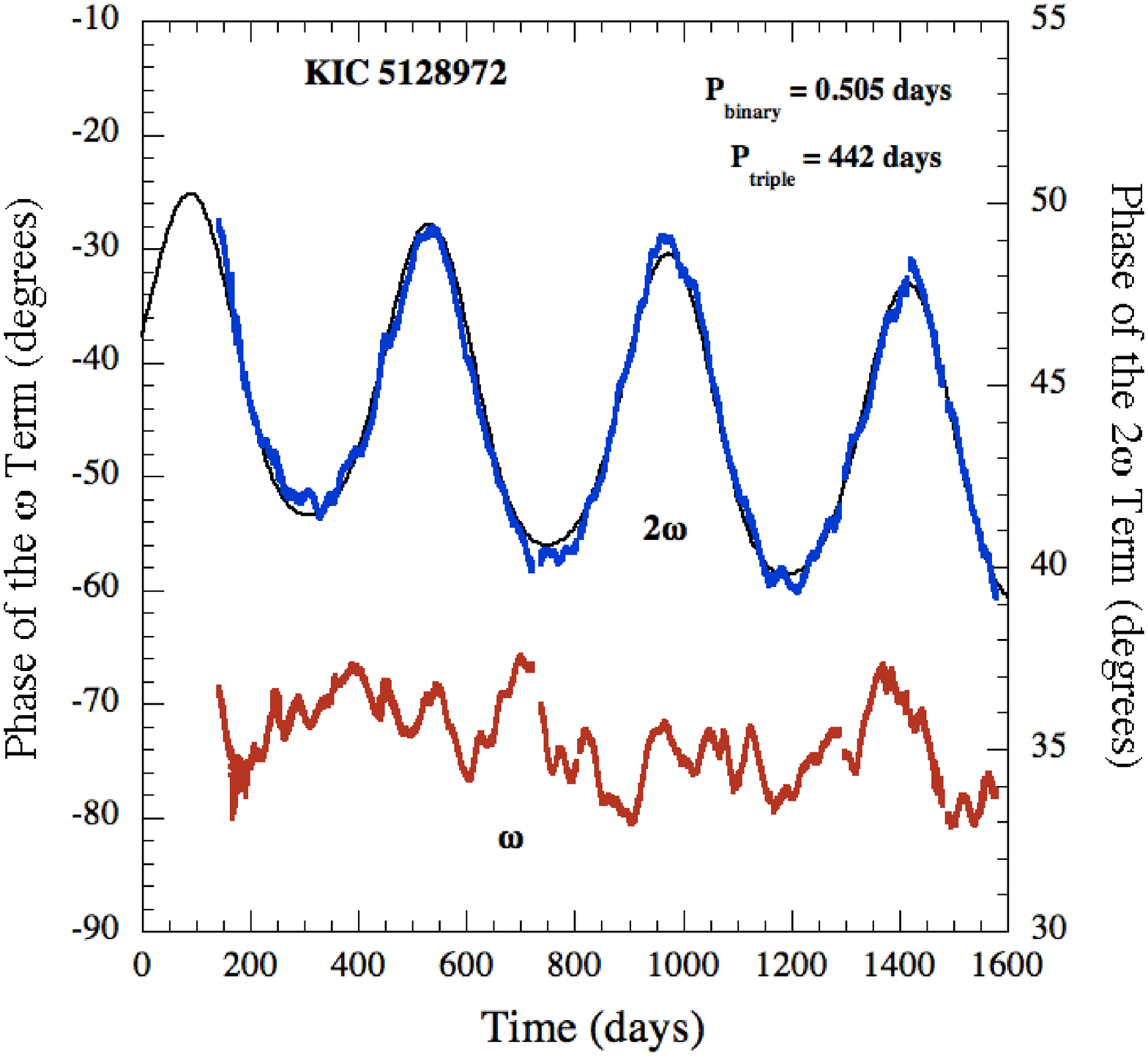}{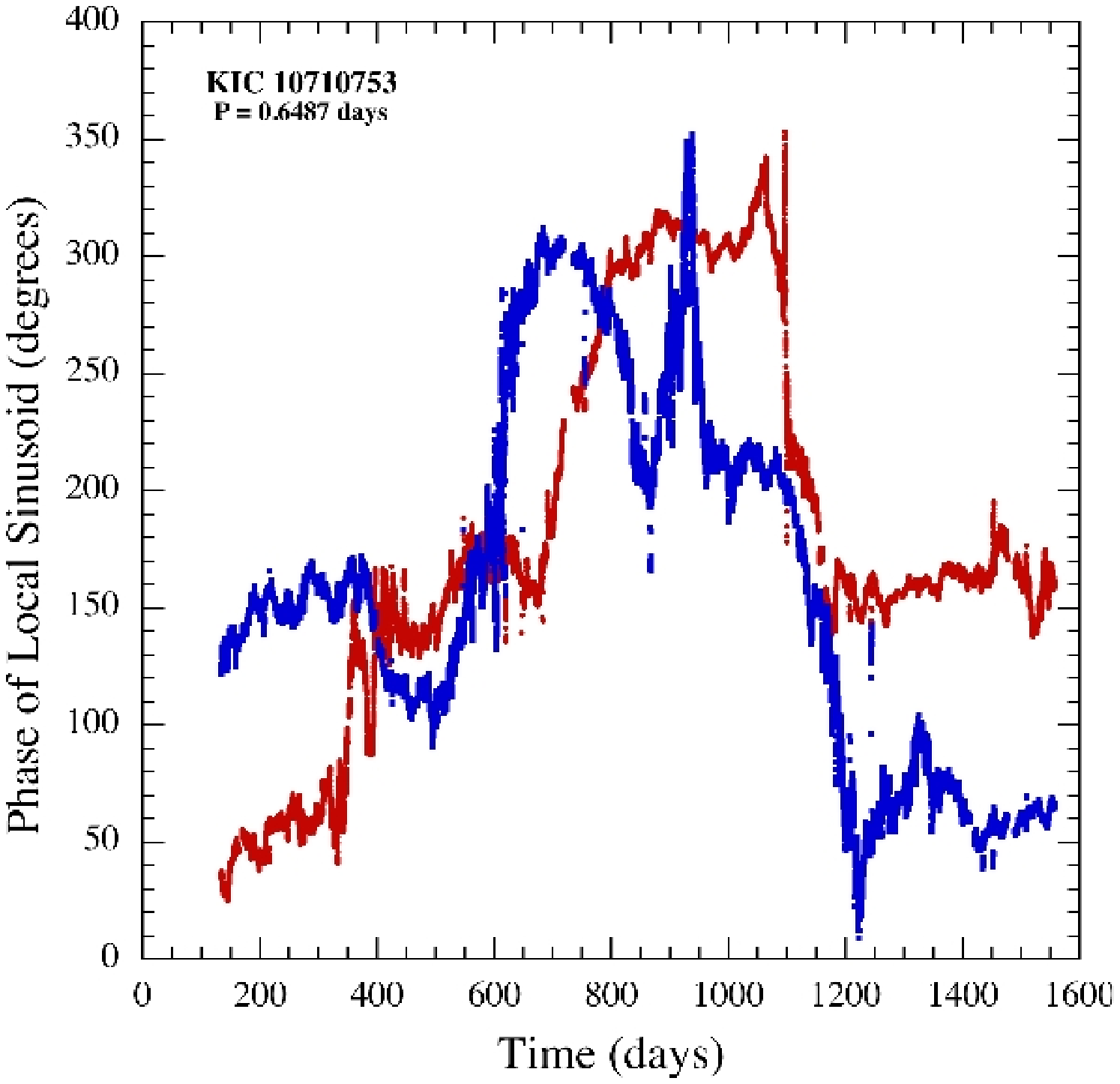}{ex_fig2}{Phase-tracking of the light curves.  \emph{Left:} Contact binary in the triple system KIC~5128972.  \emph{Right:} The spotted star KIC~10710753.  In both cases, the blue curve tracks the $\sin 2 \omega t$ term while the red curve reflects the $\sin \omega t$ term.}

\vspace{-0.2cm}

\section{Results}

Of the K-M dwarfs with short rotational periods, 54 (37 M  stars and 17 K stars) exhibit {\em two or more} independent periods. We found that differential rotation cannot be the cause of the double or multiple periods. First, this scenario would require extremely long-lasting spots (for several years), at different fixed latitudes which, based on our present knowledge of starspot behaviour,  seems unlikely. More quantitatively, measured differences between periods due to differential rotation, typically have a mean shear parameter, d$\Omega$, of about 0.07 rad/day (within a factor of $\sim$2 for cool dwarfs; see Reinhold et al. 2013, for more details). This translates to a fractional differential rotation parameter for our objects, $\alpha \simeq 0.01 P$(days), where $P$ is the stellar rotation period. From this it follows that rotation periods ($\lesssim 1$ day) which are different by more than a few percent from each other most likely do {\em not} originate from differential rotation.

\articlefiguretwo{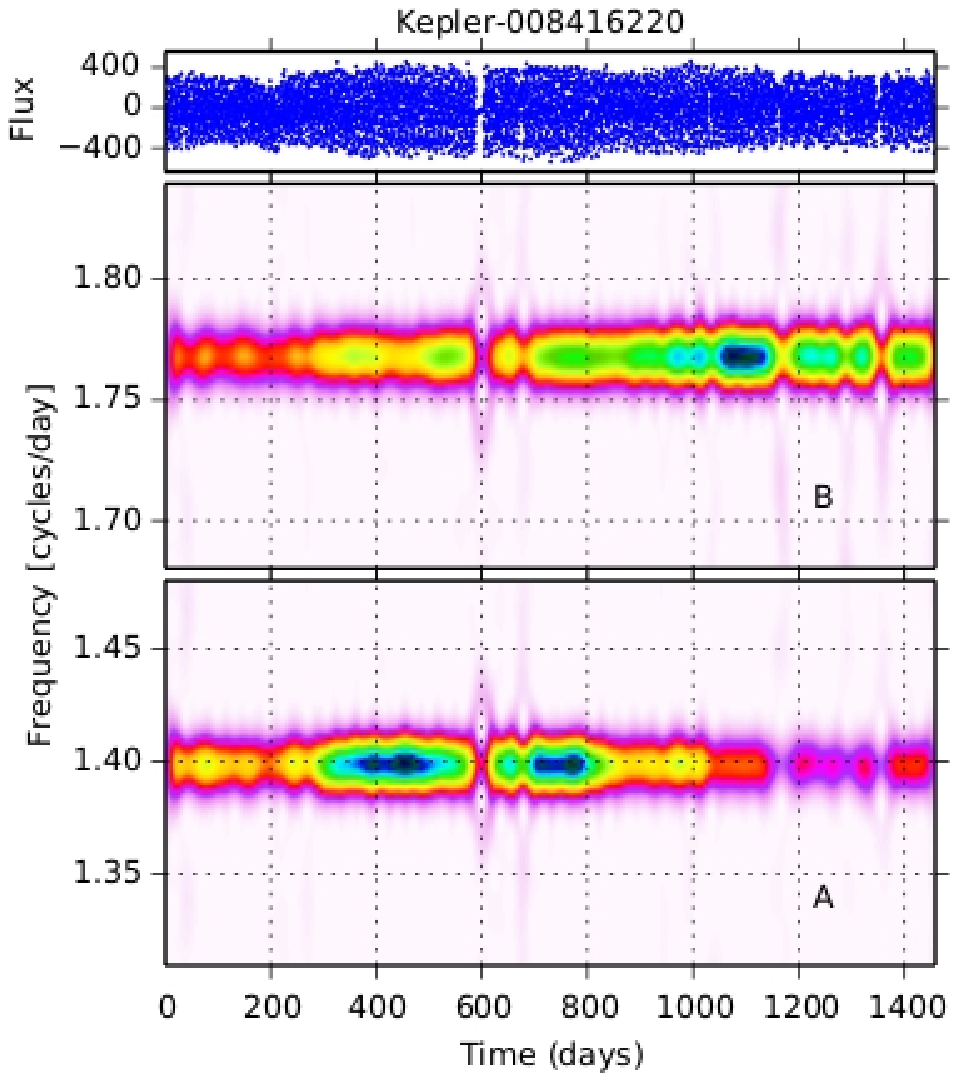}{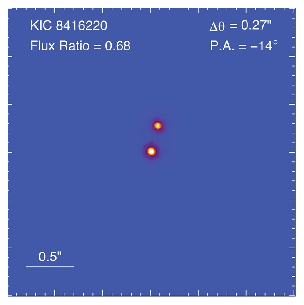}{ex_fig2}{\emph{Left:} Sonogram analysis of KIC~8416220, emphasizing the periods at 0.566 d and 0.715 d. \emph{Right:} Keck AO image of KIC~8416220 showing a close binary system (image acquired by Jonathan Swift; taken from Rappaport et al.~2014).}

The independence of the observed multiple periods is further demonstrated by the "sonogram" analysis. The observational data and the Fourier transform of KIC~8416220 are plotted in Fig.~1, while the time-frequency analysis of the same stars is shown in Fig.~3, left panel. It is evident, that the two detected periods are independent, insofar as the changes in their amplitudes are completely uncorrelated.  For all of the above reasons, it seems clear that these periods originate from different stars. 

The pixel size of Kepler is 4$''$ and the photometric aperture for each target is substantially larger.  We checked the fields of the objects having two or more short periods using the public UKIRT J-band images and, in a few cases, images obtained with adaptive optics on the Keck telescope. In the right panel of Fig.~3, the Keck image of KIC~8416220 shows dramatically there are two stars separated by only about 0.3$''$. 

In all, we find imaging evidence for the binary or multiple nature of at least 1/3 of the 54 K and M stars which exhibit two or more short rotation periods of less than 2 days.  We find separations of a few tenths to a few arcseconds. Note that a separation of 1$''$ at a distance of 200 pc implies a 200 AU physical separation, well within the expected range of bound binary separations (0.01 - $10^4$ AU). Supposing a roughly uniform stellar density in the {\em Kepler} field, we estimated the probability of interlopers at different $K_p$ magnitudes, and found that the chance alignment of the stars which we find to be multiple, is at most a few percent (see \cite{saul} for details).

\subsection{KIC~3648000, a K star with one long and two short periods}

Among the studied K dwarf stars, we found 7 objects each with one long period ($\sim$7-14 d) and two short periods ($\lesssim 1$ d). One of those is KIC~3648000 which we present here in some detail. The left panels of Fig.~4 show the high-amplitude variations due to the long period (at $\sim$12.5 d) during two well-separated time segments. The corresponding time-frequency behaviour of the periods is given in Fig.~5. In the left panel we see that the 12.5-day period is quite variable in time. The right panel indicates that the two short periods are independent of each other, having unrelated amplitude changes in time. In the right panels of Fig.~4 the two different short period variations are seen as low-amplitude waves superimposed on the 12.5-day periodicity. However, the UKIRT J-band image show no evidence of a double or multiple star system, which indicates that the separation of the anticipated binary pair (if present) must be less than 1$''$.

\articlefigure{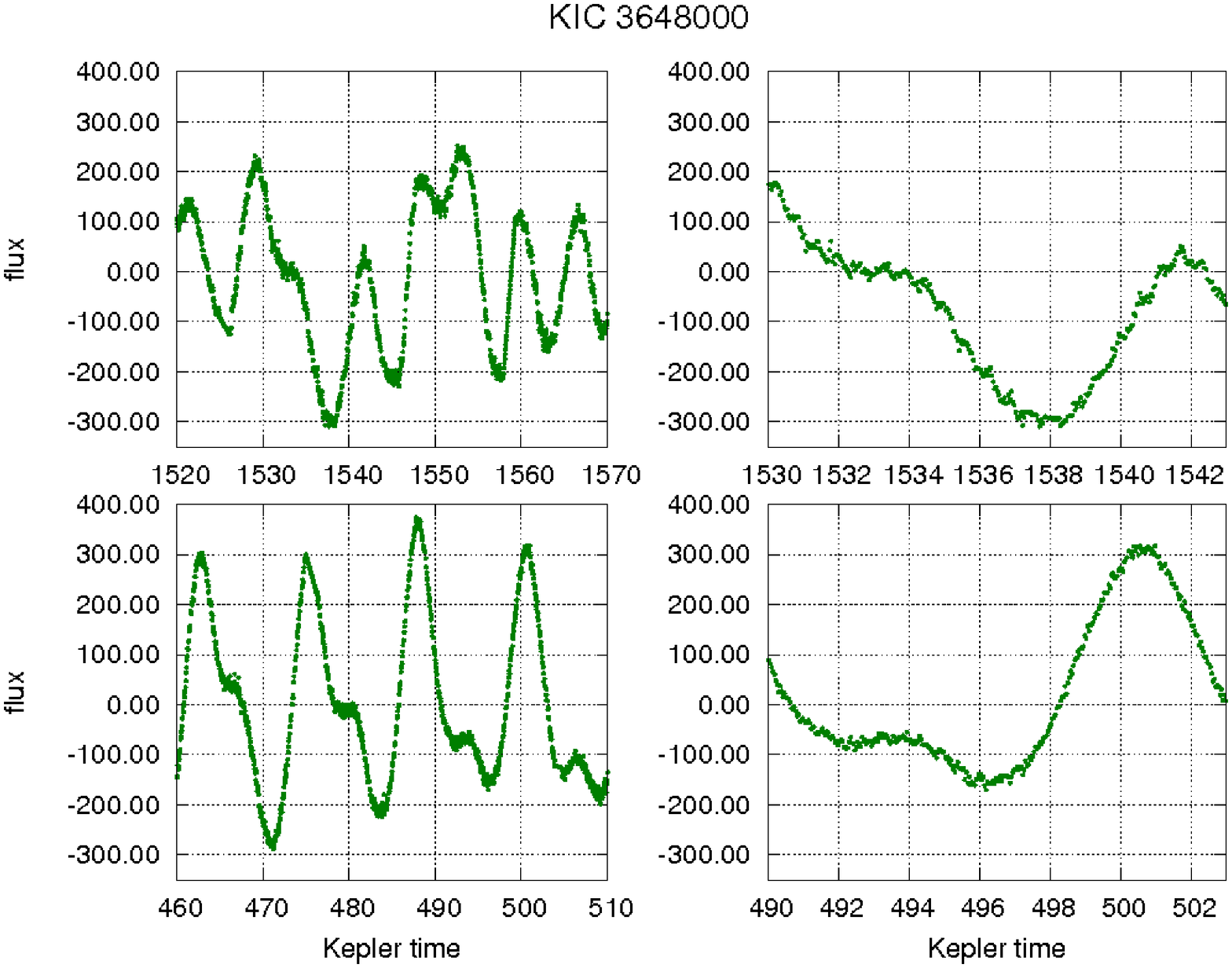}{ex_fig1}{Light curves of  KIC 3648000 during two well-separated data segments. \emph{Left two panels}: Long period (12.5 days) variation. \emph{Right:} Light variations during one rotation of the 12.5-day period; the 0.61-day period (top) and 0.54-day period (bottom) can be seen superposed with a much lower amplitude. }

\articlefiguretwo{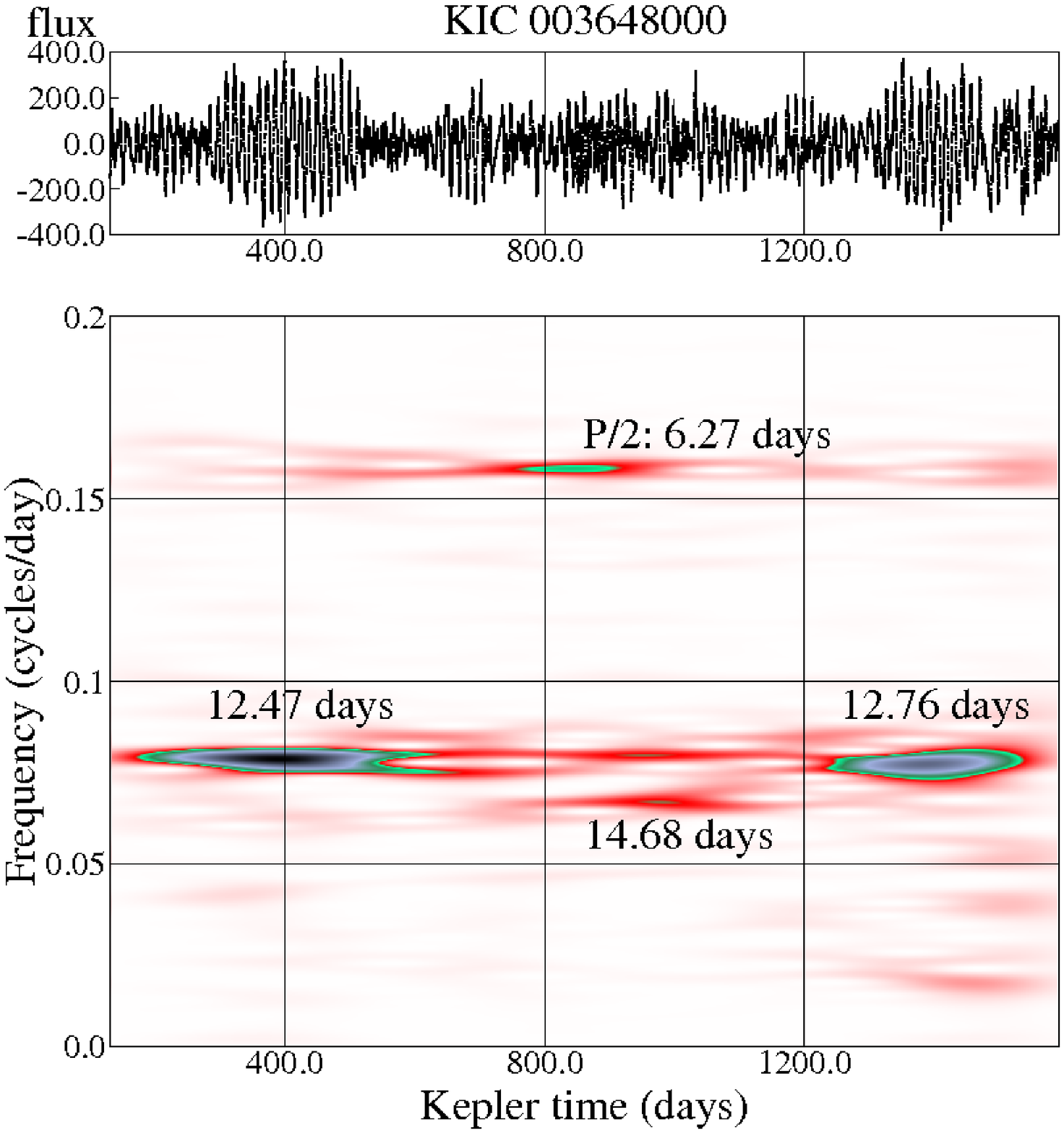}{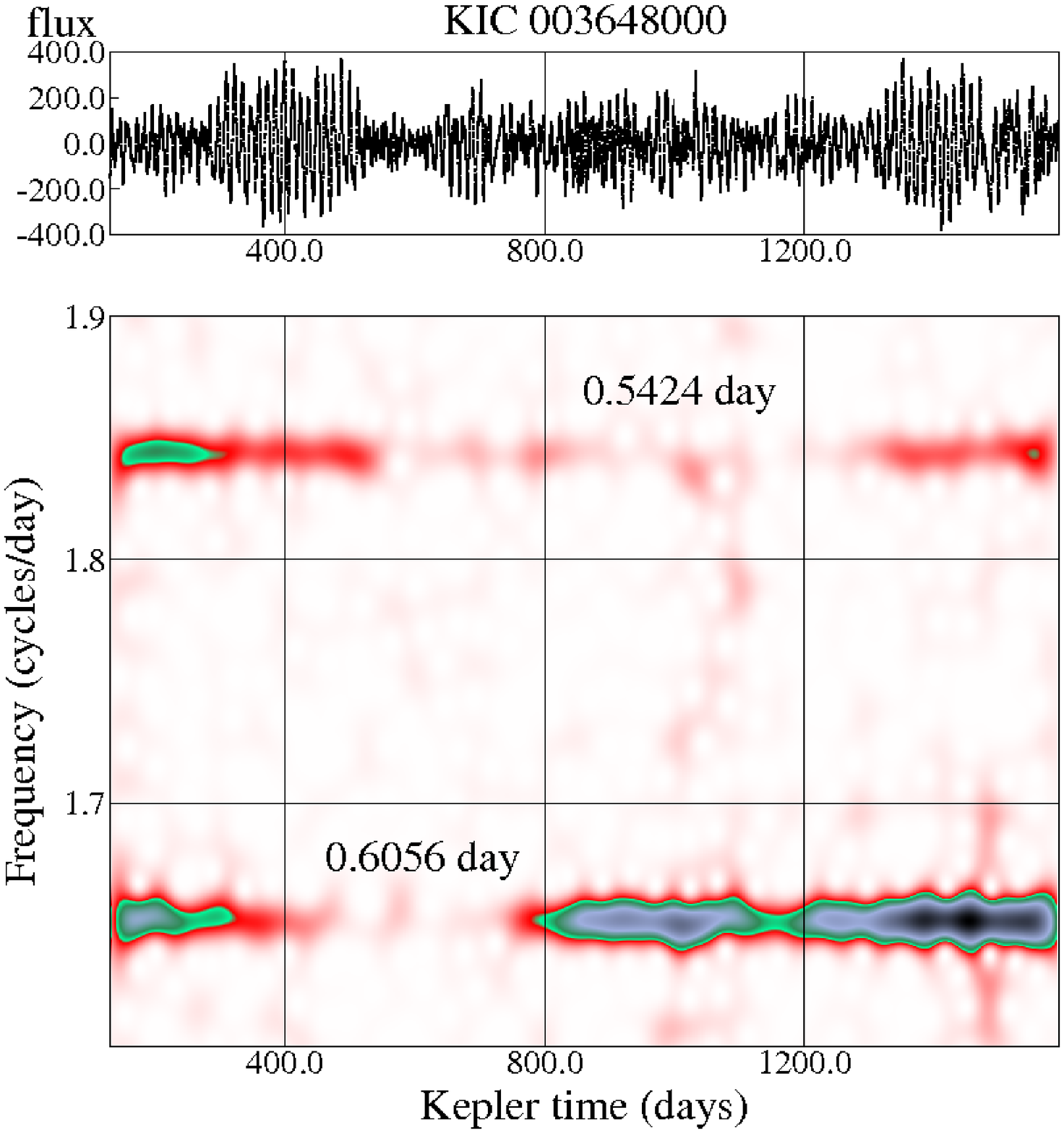}{ex_fig2}{Time-resolved Fourier transform of KIC~3648000. \emph{Left:} Variation of the 12.5-day period with time. \emph{Right:} Variation of the two short rotation periods.}

\section{Discussion}

The reason for the rapid rotation could be the youth of the stars. \citet{reinhold} found a trend that stars of spectral type F or earlier rotate faster on average than the later type ones. In late type dwarfs with convective envelopes strong magnetic fields exist which, through magnetic braking, slow down the rotation. This indicates that our rapidly rotating stars could be young.  We find that about 5\% of the M dwarfs and 2\% of the K dwarfs rotate with periods shorter than 2 days. A model of the spin evolution of such stars shows, in good accord with our finding, that M dwarfs may spend about 4\% of their lifetime having fast rotations of less than 2 days \citep{irwin}.

\section{Summary}

\noindent
$\bullet$ M and K dwarf stars with two or more short periods (differing by 
          more than a few percent) are likely bound, binary or hierarchical 
          triple systems. \\
$\bullet$ So far we have found 37 such M stars and 17 K stars with two 
          or more short periods. \\
$\bullet$ This is a novel way of discovering young, hierarchical M and K 
          star systems. \\
$\bullet$ We found that $\sim$5\% (2\%) of all M (K) stars 
	have short rotation periods of < 2 days. \\
$\bullet$ These fractions could be consistent with the amount of time that 
          an M or K star spends at short rotation periods.

\acknowledgements K.O. acknowledges support from the Hungarian research grant OTKA-109276 and from Lend\"ulet-2012 Young Researcher's Program of the Hungarian Academy of Sciences.  We made use of J-band images that were obtained with the United Kingdom Infrared Telescope (UKIRT) which is operated by the Joint Astronomy Centre on behalf of the Science and Technology Facilities Council of the U.K.


\begin{thebibliography}{}
\bibitem[Irwin et al. (2011)]{irwin}
Irwin, J., Berta, Z.K., Burke, C.J. et al. 2011, ApJ727, 56
\bibitem[Ol\'ah \& Koll\'ath (2009)]{ol_ko} Ol\'ah, K. \& Koll\'ath, Z. 2009, A\&A 501, 695
\bibitem[Rappaport et al. (2014)]{saul}
Rappaport, S., Swift, J., Levine, A. et al. 2014, ApJ 788, 114
\bibitem[Reinhold et al. (2013)]{reinhold}
Reinhold, T., Reiners, A., Basri, G. 2013, A\&A 560, 4
\end{thebibliography}
\end{document}